\documentclass[12pt]{article}
\usepackage{graphicx}

\newcommand\pubnumber{CMS CR-2018/132}
\newcommand\pubdate{\today}

\newcommand{\csh}{{\cos\theta^{*}}}
\newcommand{\sineff}{\sin^2\theta^{lept}_{eff}}

\def\napoli{Department of Physics and Astronomy\\
University of Rochester\\
Rochester, NY 14627 USA}
\def\support{\footnote{Work supported by the US Department of Energy}}

\def\Title#1{\begin{center} {\Large #1 } \end{center}}
\def\Author#1{\begin{center}{ \sc #1} \end{center}}
\def\Address#1{\begin{center}{ \it #1} \end{center}}

\newcommand\pubblock{\rightline{\begin{tabular}{l} \pubnumber\\
         \pubdate  \end{tabular}}}
\newenvironment{Abstract}{\begin{quotation}  }{\end{quotation}}
\newenvironment{Presented}{\begin{quotation} \begin{center} 
             PRESENTED AT\end{center}\bigskip 
      \begin{center}\begin{large}}{\end{large}\end{center} \end{quotation}}

\def\beq{\begin{equation}}
\def\eeq#1{\label{#1}\end{equation}}
\def\eeqn{\end{equation}}

\def\beqa{\begin{eqnarray}}
\def\eeqa#1{\label{#1}\end{eqnarray}}
\def\eeqan{\end{eqnarray}}

\let\bar=\overbar

\def\Dslash{\not{\hbox{\kern-4pt $D$}}}
\def\dslash{\not{\hbox{\kern-2pt $\del$}}}

\def\msb{{\bar{\ssstyle M \kern -1pt S}}}

\begin{document}
\begin{titlepage}
\pubblock

\vfill
\Title{Measurement of the effective weak mixing angle $sin^2\theta^{lept}_{eff}$ from the forward-backward asymmetry of Drell-Yan events at CMS}\vfill
\Author{ Arie Bodek (on behalf of the  CMS collaboration)\support}
\Address{\napoli}
\vfill
\begin{Abstract}
We report on a precision measurement of the effective weak mixing angle using the forward-backward asymmetry, $A_{FB}$, in Drell-Yan ($ee$ and $\mu\mu$) events  in pp collisions  at $\sqrt{s}=8~\mathrm{TeV}$ at the CMS  experiment at the Large Hadron Collider (LHC).  The results are compared to other hadron collider measurements published  by   ATLAS,  LHCb, CDF,
  D0,  and the Tevatron combination.
\end{Abstract}
\vfill
\begin{Presented}
13$^{the}$ International Conference on the Intersections of Particle and Nuclear Physics 
(CIPANP2018) 
\\
 Palm Springs, California  May 28-June 3, 2018
\end{Presented}
\vfill
\end{titlepage}
\def\thefootnote{\fnsymbol{footnote}}
\setcounter{footnote}{0}

\section{Introduction}

We report on a precision measurement of the leptonic effective weak mixing angle  ($\sineff$) using the forward-backward asymmetry, $A_{FB}$, in Drell-Yan ($ee$ and $\mu\mu$) events  in pp collisions  at $\sqrt{s}=8~\mathrm{TeV}$ at the CMS\cite{SMP-16-007,CMSdetector} experiment at the Large Hadron Collider (LHC).  The results are compared to other hadron collider measurements published  by   ATLAS\cite{ATLAS,ATLASdetector},  LHCb\cite{LHCb}, CDF\cite{CDF}, D0\cite{Dzero} and Tevatron\cite{Tevatron} combination.

  At leading order dilepton pairs are produced through the annihilation of a quark and antiquark to dileptons via the exchange of a  $Z$ boson or a virtual photon
 The definition of  $A_{FB}$ is based on the angle $\theta^*$ of the lepton ($\ell^-$) in the Collins-Soper  frame in the center of mass of the  dilepton system:
\begin{equation}
    A_{FB}=\frac{\sigma_{F}-\sigma_{B}}{\sigma_{F}+\sigma_{B}},
\end{equation}
where $\sigma_{F}$ and $\sigma_{B}$ are the cross sections in the forward ($\csh>0$) and backward ($\csh<0$) hemispheres, respectively. In this frame the $\theta^*$ is the angle of the $\ell^-$ direction with respect to the axis that bisects the angle between the direction of the quark and opposite direction of the anti-quark. In $pp$ collisions the direction of the quark is assumed to be in the boost direction of the dilepton pair.  In terms of  laboratory-frame energies and momenta  $\csh$  is equal to 
\begin{equation}
    \csh={\frac{2(p_1^+p_2^- - p_1^-p_2^+)}{\sqrt{M^2(M^2+P_{T}^2)} }}  \times\frac{P_z}{|P_z|},
\end{equation}
where $M$, $P_{T}$, and $P_{z}$ are the mass, transverse momentum, and longitudinal momentum, respectively, of the dilepton system, and $p_1(p_2)$ are defined in terms of the energy, $e_1 (e_2)$, and longitudinal momentum, $p_{z,1}(p_{z,2})$, of the negatively (positively) charged lepton as $p_{i}^\pm=(e_i\pm p_{{z},i})/\sqrt{2}$. 
A non-zero $A_{FB}$ in dilepton events originates from the vector and axial-vector couplings of electroweak bosons to fermions.

 The most precise  previous measurements of $\sineff$  are reported by  LEP and SLD experiments.
 However, the two most precise measurements differ  by more than 3 standard deviations.
   
 The ATLAS\cite{ATLAS} results (based on 5 fb$^{-1}$ at 7 TeV), and the LHCb\cite{LHCb} results (based on 3 fb$^{-1}$ at 7 and 8 TeV) are  published.  Measurements of $\sineff$ by  CDF\cite{CDF}, D0\cite{Dzero} and Tevatron combination\cite{Tevatron} are also published.
 Here, we focus on the most recent results from the CMS collaboration\cite{SMP-16-007}. The CMS results (based on 20 fb$^{-1}$ at 8 TeV) have been recently  accepted   for publication in the European Physics Journal C in 2018. 
 
In the CMS analysis  $\sineff$ is measured  by fitting the mass and rapidity dependence of the observed $A_{FB}$ in dilepton ($e^+e^-$ and $\mu^+\mu^-$) events. Statistical and systematic errors are reduced by using three new analysis techniques: (1) angular event weighting\cite{eventw}, (2) precise muon and electron energy calibration]\cite{momentumw} and
(3) constraining PDF errors using the $A_{FB}$ dilepton samples (Bayesian $\chi^2$ reweighting of PDF replicas)\cite{PDFw}.

\section{ Angular event weighted $A_{FB}$ at CMS}

In the Collins-Soper frame the angular distribution of dilepton events has a  (1+$\cos^2\theta^*$) term that originates from the spin 1 of the exchanged boson, a $\csh$ term  from vector-axial interference and a $(1-3\cos^2\theta^*)$ term from the transverse momentum of the interacting partons.
The angular coefficients  $A_0$ and $A_4$ are functions of dilepton mass ($M$,), transverse momentum  ($P_{T}$) and rapidity($y$)
of the dilepton pair,
\begin{equation}
\frac{1}{\sigma}\frac{{d}\sigma}{{d}\csh} = \frac{3}{8}\Big(1+\cos^2\theta^*+\frac{A_0 (M,P_T,y)}{2}(1-3\cos^2\theta^*) + A_4 (M,P_T,y)\csh\Big).
\end{equation}
In this analysis, the $A_{FB}$ values in each dilepton rapidity and mass bin are calculated using the ``angular event-weighting'' method, described in detail in Ref.~\cite{eventw}.
The technique is equivalent to measuring  $A_4$ in bins of $|\cos^2\theta^*|$, and extracting $A_{FB}$ from the average  $A_4$ for each dilepton mass bin. 

The ``angular event-weighted''  $A_{FB}$ is the same as the full phase-space $A_{FB}$, while the simple  fiducial restricted $A_{FB}$ ($A_{FB}^{restricted}$) values are smaller because of the limited acceptance at large $\csh$. Because of this feature, the event-weighted $A_{FB}$ is less sensitive to the exact modeling of the acceptance than $A_{FB}^{restricted}$. Additionally, because the event-weighted $A_{FB}$ exploits the full shape of the  $\csh$ distribution as opposed to the sign only in the  case of  $A_{FB}^{restricted}$, it also results in  a smaller statistical uncertainty in $\sineff$.

\begin{figure}[!htbp]
\centering
      \includegraphics[width=10cm, height=8cm]{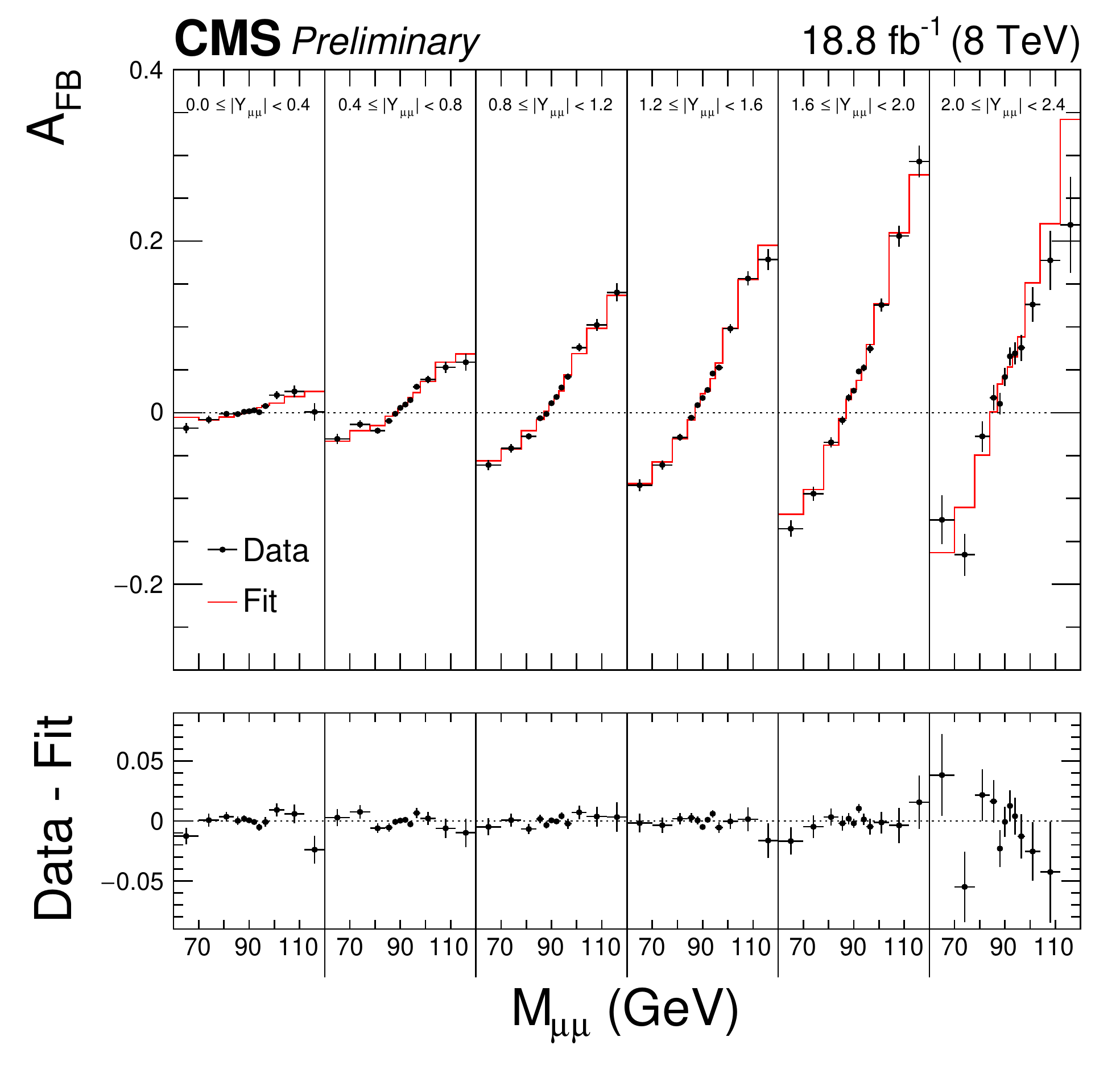}
        \includegraphics[width=10cm, height=8cm]{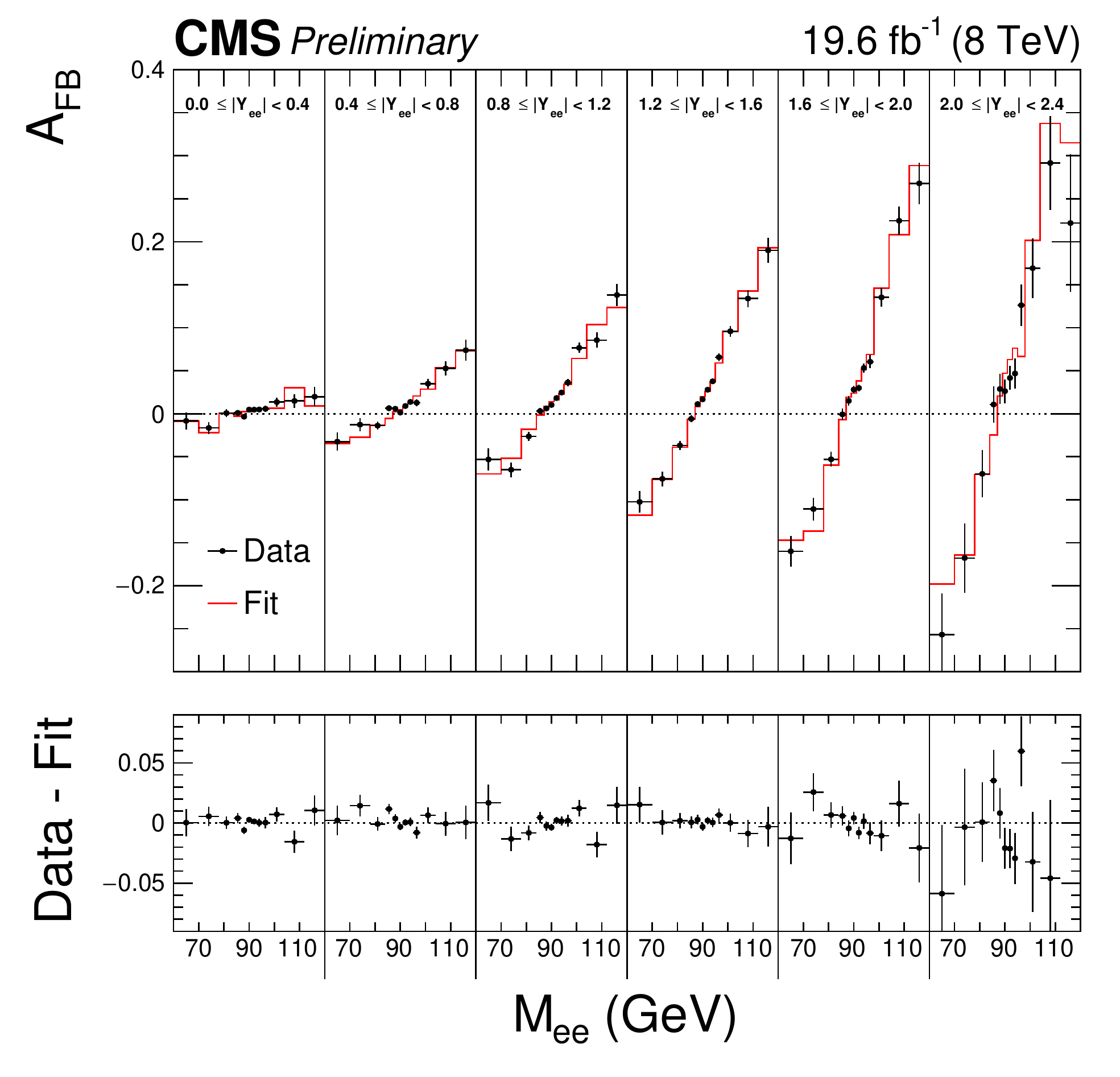}
    \caption{
	Comparison between the CMS\cite{SMP-16-007} ``angular event-weighted'' $A_{FB}$ in data and the best-fit theory prediction for  $A_{FB}$ as a function of dilepton mass for  the dimuon (top) and dielectron (bottom) channels.  The best-fit theory prediction value for  b $A_{FB}$  each bin is obtained by linear interpolation between the two neighboring best fit templates in $\sineff$. 
	The templates are based on the central PDF of the NLO NNPDF3.0 set. \label{figure:fit}
    }
\end{figure}

\section{$\sineff$ extraction at CMS}

We extract $\sineff$  by minimizing the $\chi^2$ value between the data and template $A_{FB}$ distributions in 72 dilepton mass and rapidity bins. 
The default signal templates are generated with the $\textsc{powheg}$ event generator using the NNPDF3.0 set. $\textsc{powheg}$ is interfaced with $\textsc{pythia8}$
with the CUETP8M1
 underlying event tune for parton showering and hadronization, including electromagnetic FSR. The template variations for different values of $\sineff$, renormalization and factorization scales, and PDFs are modeled using the  $\textsc{powheg}$ MC generator that provides matrix-element based event-by-event weights for each variation. To propagate these variations to the full-simulation-based templates, each event of the full-simulation sample is weighted by the ratio of $\csh$ distributions obtained with modified and default $\sineff$ configurations in each dilepton mass and rapidity bin. 

A comparison between the data and best-fit $\sineff$  template distributions is shown in Fig.~\ref{figure:fit}. 
Table~\ref{table:staterrors} summarizes the statistical uncertainty in the extracted $\sineff$ in the muon and electron channels and their combination. 
\begin{table*}[htbp]
\centering
\caption{
    Summary of the statistical uncertainties in the measurement of $\sineff$. 
    The statistical uncertainties in the lepton selection efficiency and calibration coefficients in data are included.
    \label{table:staterrors}
}
\begin{tabular}{ | l | r | r | }
\hline
channel    &  statistical uncertainty\\ 
\hline
muon		       &  0.00044  \\
electron	       &  0.00060  \\ \hline
combined	       &  0.00036  \\
\hline
\end{tabular}
\end{table*}
\begin{table*}[!htbp]
\centering
\caption{
     \label{table2}
    Summary of experimental systematic uncertainties (A)  and theory modeling uncertainties (B) in the measurement of $\sineff$ in  the dimuon (left) and dielectron (right) channels. 
    For details see ref. \cite{SMP-16-007}.
}
\begin{tabular}{ l | c | c }
\hline
Source & muons & electrons \\
\hline
MC statistics		     & 0.00015 &	0.00033 \\
Lepton momentum calibration  & 0.00008 &	0.00019 \\
Lepton selection efficiency  & 0.00005 &	0.00004 \\ 
Background subtraction	     & 0.00003 &	0.00005 \\
Pileup modeling		     & 0.00003 &	0.00002 \\
\hline
Total	 experimental  systematic uncertainties		     & 0.00018 &	0.00039 \\
\hline\hline
Model variation &  Muons &  Electrons \\ 
\hline
Dilepton $P_T$ modeling				& 0.00003 &  0.00003 \\ 
QCD $\mu_{R/F}$ scale					& 0.00011 &  0.00013 \\ 
$\textsc{powheg}$  MiNLO Z+j vs NLO Z model				& 0.00009 &  0.00009 \\
FSR model ($\textsc{photos}$ vs $\textsc{pythia}$)				& 0.00003 &  0.00005 \\ 
UE tune							& 0.00003 &  0.00004 \\ 
Electroweak ($\sin^2\theta^{{lept}}_{{eff}} - \sin^2\theta^{{u, d}}_{{eff}}$ ) & 0.00001 &  0.00001 \\ 
\hline 
Total theory modeling uncertainties						& 0.00015 &  0.00017 \\ 
\hline
\end{tabular}
\end{table*}

The systematic and theory modeling uncertainties are summarized in Table ~\ref{table2}. A detailed discussion of the  experimental systematic and theory modeling uncertainties is given in Ref. \cite{SMP-16-007}.  
\section{Constraining PDFs with $A_{FB}$ data and Bayesian $\chi^2$ PDF reweighting at CMS}
The observed  $A_{FB}$ values depend on the size of the dilution effect, as well as on the relative contributions from u and d valence quarks to the total dilepton production cross section. Therefore, the PDF uncertainties translate into sizable variations in the observed  $A_{FB}$ values.
However, changes in PDFs affect the $A_{FB}$($M_{\ell\ell}$, $Y_{\ell\ell}$) distribution in a different way from changes in $\sineff$. 

Changes in PDFs result in changes in $A_{FB}$  in regions where the absolute values of $A_{FB}$ is large, i.e. at high and low dilepton masses. In contrast,
 the effect of changes in $\sineff$ are largest near the Z-peak and are significantly smaller at high and low masses. Because of this behavior.
 we can apply the Bayesian $\chi^2$ reweighting method to 100 NNPDF3.0 PDF replicas to  constrain the PDFs~\cite {PDFw, Giele, Sato} and  reduce the PDF errors in the extracted value of $\sineff$.

\begin{figure}[ht]
\centering
    \includegraphics[width=5.2cm]{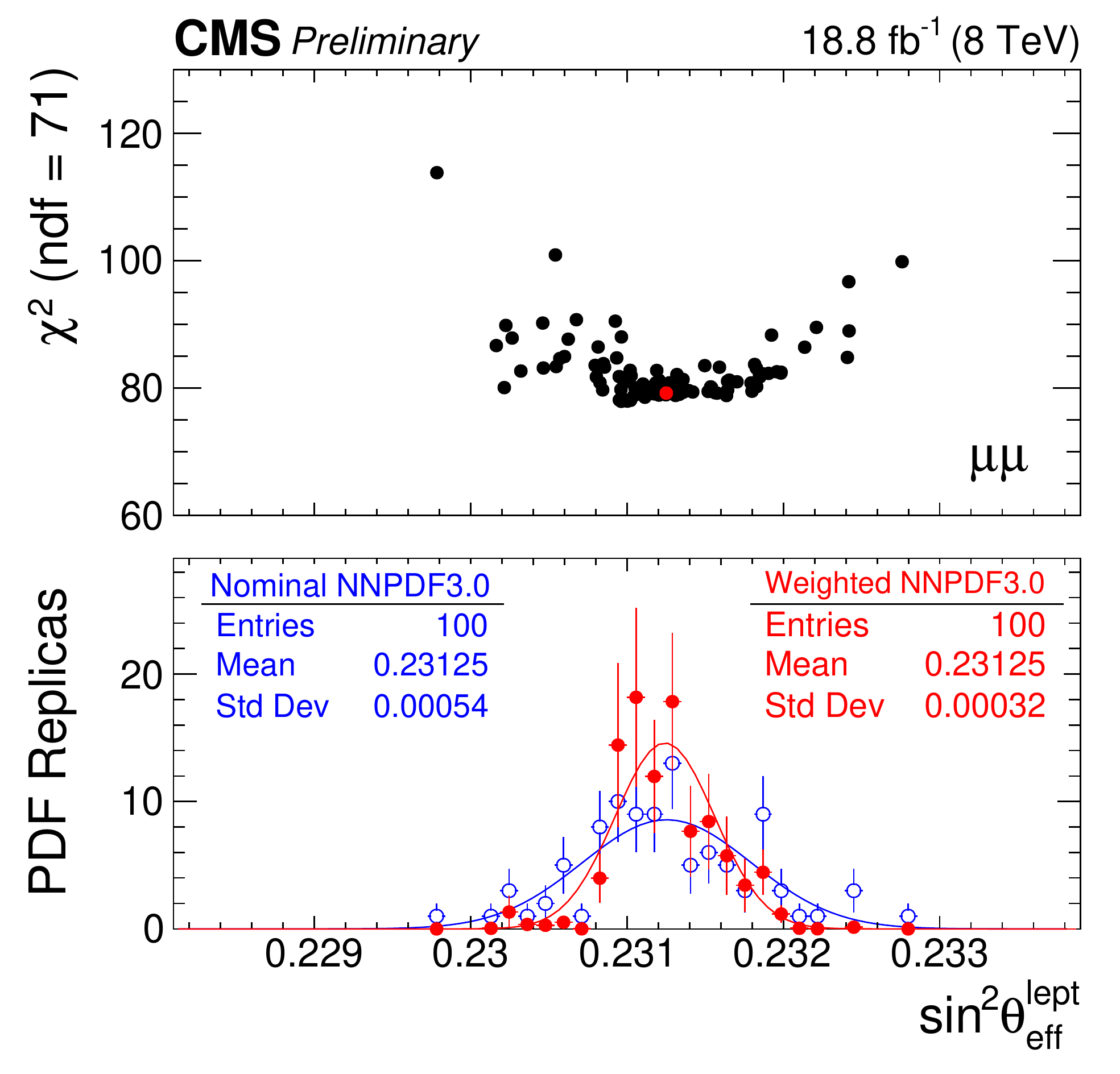}
    \includegraphics[width=5.2cm]{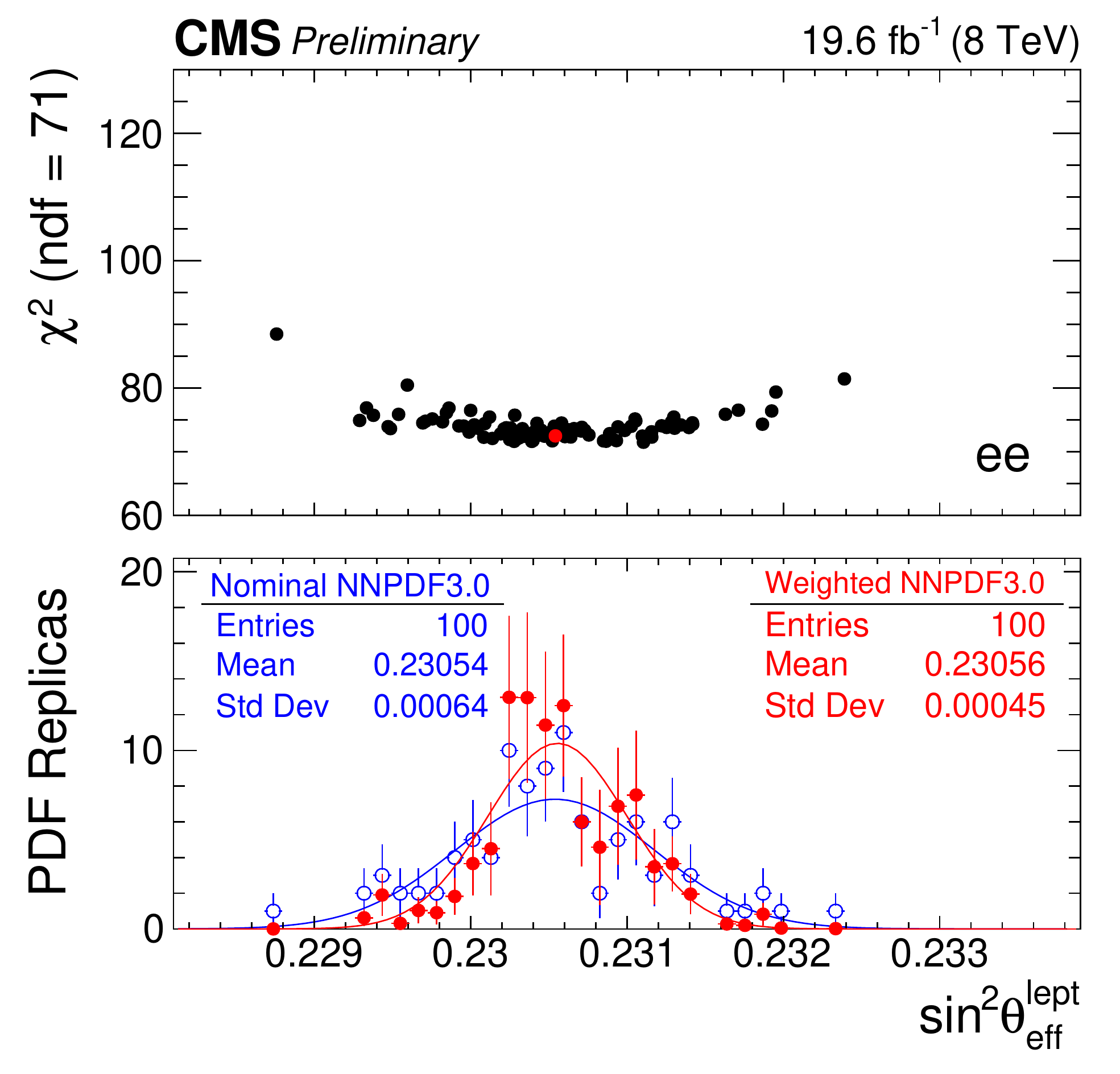} \\
    \includegraphics[width=5.2cm]{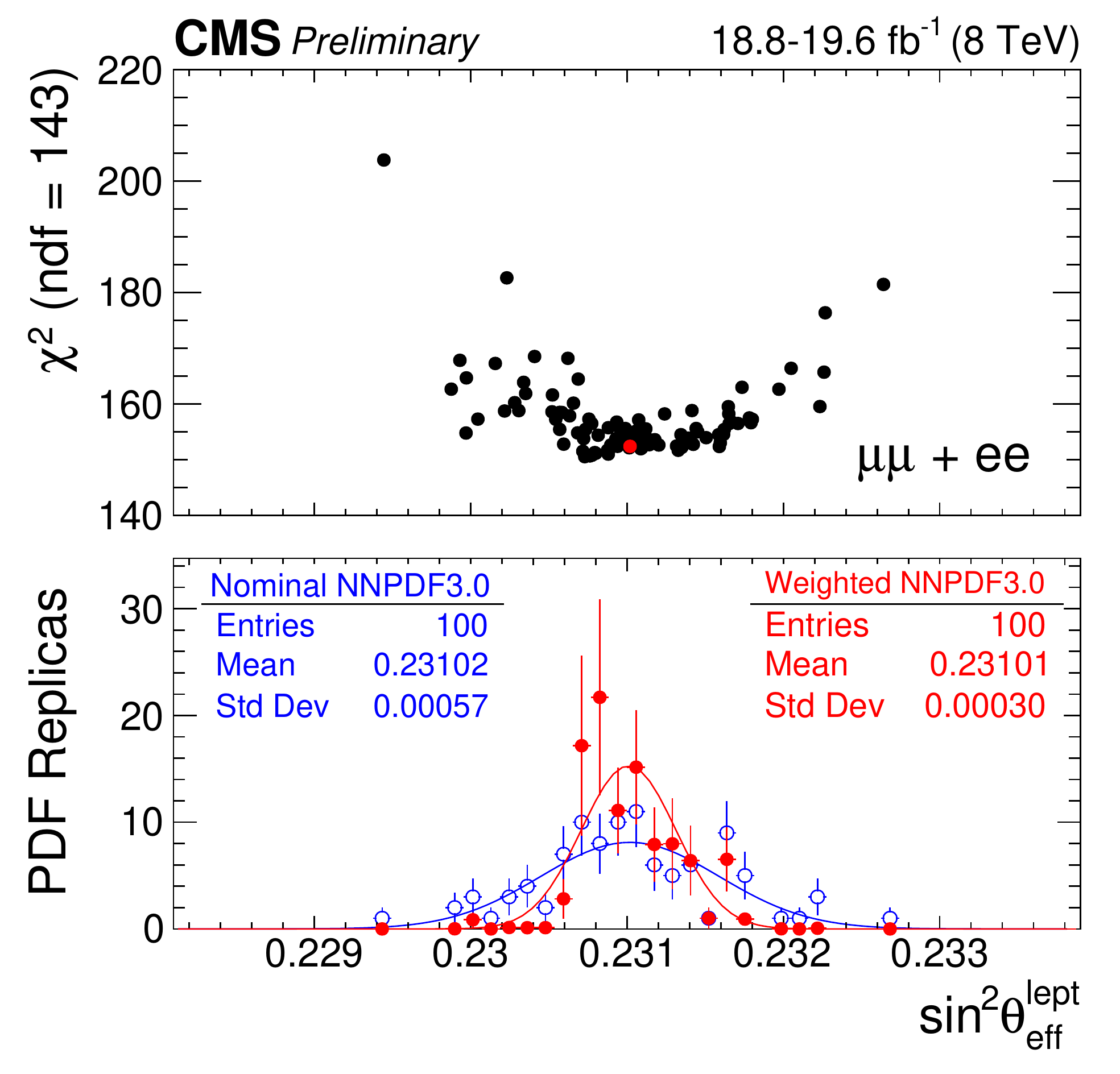}
    \caption{
	The top panel of each figure shows the $\chi^2_{min}$ vs best-fit 
	$\sineff$ distribution for 100 NNPDF replicas in muon channel (top left), 
	electron channel (top right), and their combination (bottom). 
	The corresponding bottom panels show the best-fit $\sineff$ distribution 
	over the nominal (blue) and weighted (red) PDF replicas\cite{SMP-16-007}.  
	\label{figure:comination:scatter}
    }
\end{figure}

Fig.~\ref{figure:comination:scatter} shows the distribution of the $\chi^2_{min}$ vs best-fit $\sineff$  
value for  the 100 NNPDF3.0 replicas for the $ee$, $\mu\mu$ samples and  combined $ee$+$\mu\mu$ samples. In these plots, all sources of the statistical and experimental systematic uncertainties are included into the $72\times72$ covariance matrices for both data and template $A_{FB}$ distributions.
As illustrated in these figures, the extreme PDF replicas from either side are disfavored by both the dimuon and dielectron data. For each of the NNPDF replica, the electron and muon results are combined using their respective best-fit $\chi^2$ values, $\sineff$ and statistical and experimental systematic uncertainties. 

 %
The extracted $\sineff$ in the electron and muon decay channels and their combination with and without constraining the PDF uncertainties are shown in Table~\ref{table:combination}. 
 After Bayesian $\chi^2$ reweighting by  $w_i = [{e^{-\frac{\chi^2_{{min},i}}{2}}}]/[{\frac{1}{N}\sum_{i=1}^N e^{-\frac{\chi^2_{{min},i}}{2}}}]$,
  the PDF uncertainties are  reduced by about a factor of 2. It should be noted that the Bayesian $\chi^2$ reweighting technique works well if the PDF replicas span the optimal value on both sides.  Additionally, the effective number of replicas after $\chi^2$ reweighting, $n_{{eff}}=N^2/\sum_{i=1}^{N}w_i^2$, should also be large enough to give a reasonable estimate of the average and the standard deviation. The number of effective replicas after the $\chi^2$ reweighting is $n_{{eff}}=41$. As a cross check we also perform the analysis with the corresponding 1000-replica NNPDF set in the dimuon channel and find good agreement between the two results.
\begin{table}[!htbp]
\centering
\caption{
    Central value and PDF uncertainty of the measured $\sineff$ in the muon and electron channels and their combination 
    with and without constraining PDFs using Bayesian $\chi^2$ reweighting.
}
\label{table:combination}
\begin{tabular}{ l | c | c }
\hline
Channel		   &  without constraining PDFs	    & with constraining PDFs \\ \hline 
Muon               &  $0.23125\pm0.00054$ & $0.23125\pm0.00032$  \\
Electron           &  $0.23054\pm0.00064$ & $0.23056\pm0.00045$  \\
Combined           &  $0.23102\pm0.00057$ & $0.23101\pm0.00030$  \\
\hline
\end{tabular}
\end{table}
\vspace{-0.1in}
\section{Summary}
We report on the extraction\cite{SMP-16-007} of  $\sineff$ from the measurements of the mass and rapidity dependence of $A_{FB}$ in Drell-Yan $ee$ and $\mu\mu$ 
events in pp collisions at $\sqrt{s}=8~\mathrm{TeV}$ at CMS.  
With larger samples and new analysis techniques (including precise lepton momentum calibration, angular event weighting, and additional PDF constraints from  Bayesian $\chi^2$ reweighting),  the statistical and systematic uncertainties are reduced by a factor of two compared to previous measurements at the LHC\cite{ATLAS,LHCb}. 
The combined result from the dielectron and dimuon channels is:
\begin{eqnarray} \nonumber
	\sineff~&=&
	0.23101\pm
            0.00036({stat})\pm
	    0.00018({syst})\pm
	    0.00016({th})\pm
	    0.00030({pdf}), {\rm or} \nonumber \\
	\sineff &=&0.23101\pm0.00052 ~(\rm CMS~ 8 ~TeV). \nonumber
\end{eqnarray}
\begin{figure}[ht]
    \includegraphics[width=10cm, height=9cm]{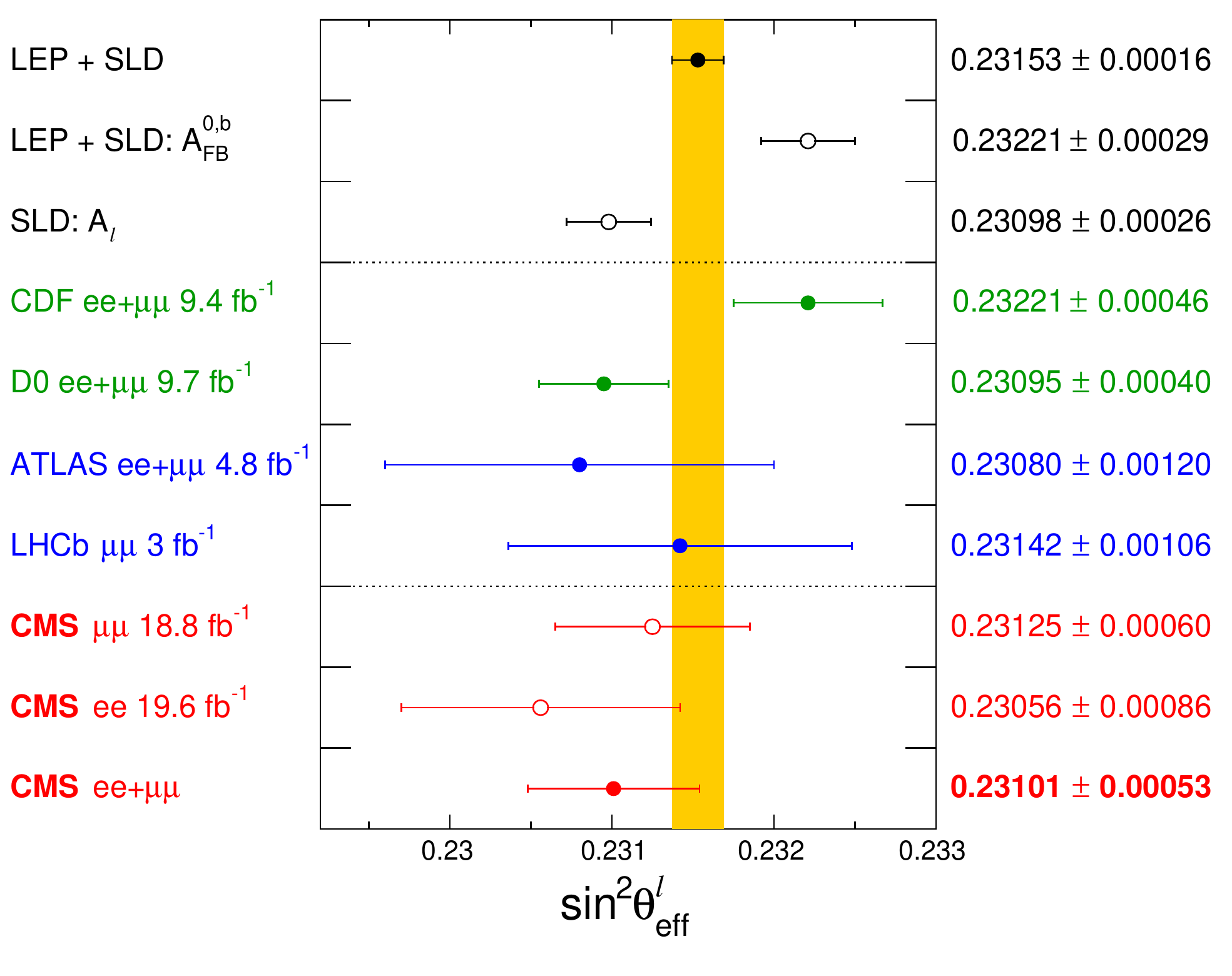}
    \centering
    \caption{
	Comparisons of the measured CMS\cite{SMP-16-007} $\sineff$ in the $\mu^+\mu^-$ and $e^+e^-$ channels and their combination with
	 LEP/SLC, LHC\cite{ATLAS,LHCb} and  Tevatron\cite{CDF,Dzero,Tevatron} measurements.  
	The shaded band corresponds to the combination of LEP and SLC.
	\label{figure:result}
    }
\end{figure}
\vspace{-0.05in}
Comparisons of the measured CMS\cite{SMP-16-007} $\sineff$ in the muon and electron channels and their combination with
	previous LEP/SLC, LHC\cite{ATLAS,LHCb} and  Tevatron\cite{CDF,Dzero,Tevatron} measurements are shown in Figure~\ref{figure:result}. 
	The shaded band corresponds to the combination of the LEP and SLC measurements.    The results are consistent with the most precise LEP and SLD measurements. 
\subsection{Preliminary results presented by ATLAS at ICHEP 2018}
The above represents the status of measurements as presented at the  time of the  CIPANP conference in May 2018.   More recent preliminary results were presented by
ATLAS\cite{ATLASnew} at ICHEP in July of 2018.  The new ATLAS results are based on 8 TeV data, and include results from the forward electromagnetic
detector.  The ATLAS results incorporate constraining PDFs by
constraining Hessian PDF nuisance parameters from the mass and rapidity dependence of $A_{FB}$.   Constraining Hessian PDF nuisance parameters is equivalent\cite{SMP-16-007}  to constraining PDF by re-weighting Bayssian PDF replicas.  The ATLAS ICHEP 2018 preliminary results are:
\begin{eqnarray}
	\sineff &=&0.23140\pm
            0.00021({stat})\pm
	    0.00016({syst})\pm
	    0.00024({pdf}), {\rm or} \nonumber \\
	 \sineff &=& 0.23140\pm0.00036 ~\rm(ATLAS~ 8 ~TeV~ICHEP~2018~preliminary).\nonumber
\end{eqnarray}


%
%


\begin{thebibliography}{99}
%
%
%
                         \bibitem{SMP-16-007} CMS Collaboration, ``Measurement of the weak mixing angle with the forward-backward asymmetry of Drell-Yan events at 8 TeV'', 
CMS-PAS-SMP-16-007 (2017), https://cds.cern.ch/record/2273392?ln=en.
 \bibitem{CMSdetector}  CMS Collaboration, JINST 3 S08004 (2008).

\bibitem{ATLAS} ATLAS Collaboration, ``{Measurement of the forward-backward asymmetry of
            electron and muon pair-production in $pp$ collisions at   $\sqrt{s}$ = 7 TeV with the ATLAS detector}'', 404 JHEP 09 (2015) 049.   
 \bibitem{ATLASdetector}  ATLAS Collaboration, JINST 3 S08003 (2008).

                        
\bibitem{LHCb} LHCb Collaboration,  ``{Measurement of the forward-backward asymmetry in
                        $Z/\gamma^{\ast} \rightarrow \mu^{+}\mu^{-}$ decays and     determination of the effective weak mixing angle}",
                        JHEP 11 407 (2015) 190.
    
\bibitem{CDF}  T. Aaltonen et al. (CDF Collaboration), Phys. Rev. D93, 112016  (2016).
\bibitem{Dzero}   V. M.  Abazov et al. (D0 Collaboration), Phys. Rev. Lett. 120, 241802 (2018).
\bibitem{Tevatron} T. Aaltonen et al. (CDF Collaboration, D0 Collaboration),  Phys. Rev. D97, 112007 (2018). 
\bibitem{eventw}  A. Bodek,  ``A simple event weighting technique for optimizing the measurement of the
forward-backward asymmetry of Drell-Yan dilepton pairs", Eur. Phys. J. C 67 (2010) 321.
\bibitem{momentumw} A. Bodek et al.,   ``Muon Momentum Scale Corrections for Hadron Collider
Experiments", Eur. Phys. J. C 72 (2012) 2194.
\bibitem{PDFw} A. Bodek, J. Han, A. Khukhunaishvili, and W. Sakumoto,  ``Using Drell-Yan
forward-backward asymmetry to reduce PDF uncertainties in the measurement of
electroweak parameters", Eur. Phys. J. C 76 (2016) 115.
\bibitem{Giele} W. T. Giele and S. Keller,  ``Implications of hadron collider observables on parton
distribution functions",  Phys. Rev. D 58 (1998) 094023.
\bibitem{Sato}  N. Sato, J. F. Owens, and H. Prosper,  ``Bayesian Reweighting for Global Fits",  Phys. Rev.
D 89 (2014) 114020.
\bibitem{ATLASnew} ATLAS Collaboration, ``Measurement of the effective leptonic weak mixing angle using electron and muon pairs from Z-boson decay in the ATLAS experiment at $\sqrt{s}$ = 8  TeV",
ATLAS-CONF-2018-037,  https://cds.cern.ch/record/2630340.
\end{thebibliography}
\end{document}